\documentstyle[preprint,epsfig,aps]{revtex}

\begin{document}

\centerline{\bf Calculation Methods for 
Radio Pulses from High Energy Showers}

\centerline{J. Alvarez-Mu\~niz} 

\centerline{\it Department of Physics, University of Wisconsin,
Madison, WI 53706, USA}

\centerline{R.A. V\'azquez and E. Zas}

\centerline{\it Departamento de F\'\i sica de Part\'\i culas, Universidade de
Santiago}
\centerline{\it E-15706 Santiago de Compostela, Spain}

\begin{abstract}
We present an approximation for the numerical calculation of \v Cerenkov 
radio pulses in the Fraunhofer limit from very high energy showers 
in dense media. 
We compare it to full Montecarlo simulations in ice 
studying its range of applicability and show how it can 
be extended with a simple algorithm. 
The approximation reproduces well the angular distribution 
of the pulse around the \v Cerenkov direction. 
An improved parameterization for the frequency spectrum in the 
\v Cerenkov direction is given for phenomenolgical applications. 
We extend the method  
to study the pulses produced by showers at distances at which the 
Fraunhofer limit does not apply, and give the ranges of distances and  
frequencies in which Fraunhofer approximation is good enough for 
interpreting future experimental data. Our results are relevant for 
the detection of very high energy neutrinos with this technique.  
\end{abstract}

{\bf PACS number(s):} 96.40.Pq; 96.40.Tv; 95.85.Bh; 13.15.+g 

{\bf Keywords:} \v Cerenkov radiation, LPM effect, 
Electromagnetic and hadronic showers, Neutrino detection.


\section{Introduction}

The confirmed detection of 
cosmic rays above the Greisen-Zatsepin-Kuz'min cutoff 
gives confidence in the existence of neutrinos of energies 
reaching the EeV scale and above. 
Such neutrinos are expected both in models in which the 
protons are accelerated to the highest energies \cite{nufluxes}, 
such as in Active 
Galactic Nuclei \cite{mannheim} or Gamma Ray Bursts \cite{bahcall} 
and in ``top bottom" scenarios \cite{Berezinsky,sigl} 
in which cosmic rays are basically produced through quark 
fragmentation in events such as the decay of long lived 
heavy relic particles \cite{Sarkar} 
or the annihilation of topological defects \cite{TD}. 
If the highest energy component of the cosmic rays are protons,
as suggested by increasing experimental evidence \cite{ave00,vazquez,yakutsk}, 
they are expected to produce neutrinos in their interactions with the 
cosmic microwave background \cite{stecker}. 
Neutrino detection would provide extremely valuable information on
fundamental questions, both in astrophysics, such as the origin 
of the highest energy cosmic rays and in particle physics. 

Detecting high energy neutrinos may be a 
reality in the immediate future as many efforts are 
being made to develop large scale \v Cerenkov detectors 
under water or ice \cite{AABN}, designed to challenge
the low neutrino cross section exploiting the 
long range of the high energy muons produced in charged 
current muon neutrino interactions.  
For EeV neutrinos 
these detectors are also capable of detecting light from 
high energy showers produced by neutrinos of any flavor 
in both neutral and charged current interactions, but 
the effective acceptance of the detector is reduced because 
the shower must be produced very close or within the 
instrumented volume.  

It has been known for long that the development of showers in 
dense media produces an excess charge which generates 
a coherent \v Cerenkov pulse in the radiowave frequency when it 
propagates through the medium \cite{askaryan}. 
The detection of these pulses provides a possible alternative 
to neutrino detection particularly appropriate for very high 
energies \cite{ralston,provorov,price}  
because the signal scales with the square of the primary 
energy \cite{markov,zas92}. The method is attractive because 
of the good transmission properties of large natural volumes of 
ice and sand and because much information about the charge 
distribution in the shower is preserved in the frequency and 
angular distribution of the pulses. 
This last property can be used to extract information about 
shower energy and neutrino flavor \cite{alz99}. 
The technique faces a number of technical difficulties however 
\cite{jelley} and several attempts are currently being made to 
test the theoretical predictions \cite{Argonne} and to study 
the feasibility of the technique in Antarctic ice \cite{RICE}. 

Theoretical calculations are also difficult because a 
complete interference calculation calls for simulations capable 
of following electrons and positrons to the \v Cerenkov threshold 
($\sim 100~$keV). For high energy showers this is unfortunately 
out of question because of the large number of particles involved 
and approximations have been specifically deviced to study the 
radio emission of high energy showers in ice.
The calculation of radio pulses from EeV showers has been
possible in the {\sl one dimensional} (1-D) {\sl approximation}
which consists on neglecting both the lateral distribution
and the subluminal velocity of shower particles \cite{alz99,alz97,alz98}.
All the calculations of radio pulses have been made so far in the 
Fraunhofer limit. In this limit the dependence of the electric 
field on distance to shower is trivial and the characterization 
of the angular distribution of the radio pulse at a given frequency 
is effectively only dependent on one variable, namely the angle between 
the shower axis and the observation direction what simplifies the 
simulations \cite{zas92}. 
Clearly Fresnel type interference will take place if the showers are 
close enough to the detectors, but the calculation of these effects 
becomes even more time consuming. 

In this paper we firstly give a brief introduction to coherent 
radio emission in Section II (fuller details can be found in Refs.~
\cite{zas92,Allan}) accounting for the approximations made. 
In Section III we make extensive tests and explore 
the validity of the 1-D approximation in the Fraunhofer limit 
by direct comparison with complete simulations, and 
we discuss the approximation pointing out the 
connections between the radio emission and shower fluctuations, 
what gives new and useful insight into the radioemission processes. 
In Section IV we use the 1-D approximation without taking the Fraunhofer 
limit to study the radiopulse as a function of the distance to observation 
point. 
In Section V we summarize and conclude, commenting on the implications 
of our results for neutrino detection. 

\section{\v Cerenkov Radio Pulses}

When a charged particle travels through a dielectric medium
of refraction index $n$ with speed $\beta c$ greater than
the phase velocity of light in that medium ($c/n$),
then \v Cerenkov radiation is emitted in a frequency band over which
the $~\beta n > 1$ condition is satisfied without large absorbtion.
The calculation of the \v Cerenkov electric field associated
to the particle is a problem of classical electromagnetism
that has been addressed elsewhere \cite{jackson}.  
Solving the inhomogeneous Maxwell's equations in the transverse gauge,
it is easy to obtain the Fourier components of the 
electric field produced by a current density
${\vec J} ({\vec x'},t')$: 

\begin{equation}
{\vec E}({\vec x},\omega)=
{e \mu_{\rm r}~\over 2 \pi \epsilon_0 {\rm c}^2}~i \omega
\int\int\int\int dt'd^3 {\vec x'}~ {e^{i\omega t'+
ik\vert {\vec x}-{\vec x'}\vert}\over \vert {\vec x}-{\vec x'}\vert}
{\vec J}_\perp ({\vec x'},t')
\label{efield}
\end{equation}
where ${\vec J}_\perp ({\vec x'},t')$ is the component
of the current transverse to the direction of observation ${\vec x}$. 
Also $\nu$ ($\omega$) is the frequency (angular frequency), 
$k$ is the modulus of the wave vector 
$\vec k$, $\mu _r$ is the relative permeability of the medium and 
$\epsilon _0$ and $c$ is the permittivity and velocity of light in the 
vacuum. 

A powerful approach to the simulation problem can be obtained neglecting
the lateral distributions in shower particles and assuming all particles
move at constant speed $c$ in one dimension. 
We obtain a useful compact expression relating 
the charge distribution of the shower and its associated electric field. 
Crude as it may look, this
approximation (1-D approximation in brief) will be
shown to give very good results particularly around the \v Cerenkov angle
and it has allowed the possibility of establishing the radioemission from
EeV showers \cite{alz97,alz98}. The method naturally relates different
features of shower development to the spectrum and angular distribution
of the radio emission in an interesting way, giving insight into the
complexity of the calculated angular pulses.
For simplicity we are going to take ${\vec x'}={\vec z}~'=z'{\hat n_{z'}}$
where ${\hat n_{z'}}$ is a unitary vector along the shower axis. 
The current associated to the shower development in this approximation
is then given by:

\begin{equation}
{\vec J}_\perp (\vec z~',t')=Q(z')~
{\vec c}_\perp ~\delta^3 (\vec z~'-{\vec c}t')
\label{current}
\end{equation}
where $Q(z')$ is the longitudinal development of the excess charge
in the shower. The substitution of this current into 
Eq.~\ref{efield} leads to:

\begin{equation}
{\vec E}({\vec x},\omega)=
{e \mu_{\rm r}~\over 2 \pi \epsilon_0 {\rm c}^2}~i \omega
~\sin\theta~{\hat n_\perp}~\int dz'~ Q(z')~{e^{i{\omega\over c} z'+
ik\vert {\vec x}-z'{\hat n_{z'}}\vert}\over \vert {\vec x}-z'{\hat n_{z'}} \vert}
\label{fresnel}
\end{equation}
where $\theta$ is the angle between the shower axis and the direction 
of observation ${\vec x}$ and ${\hat n_\perp}$ is a unitary vector
perpendicular to ${\vec x}$.

We can use this expression to
obtain the \v Cerenkov electric field emitted by a particle
shower propagating along a medium. Eq.~\ref{fresnel} accounts for 
the correct phase factors and distances for showers that are close 
to the observer (Fresnel region).
In the Fraunhofer limit the phase factor in Eq.~\ref{fresnel}
can be approximated by $ik\vert {\vec x}-{\vec z}~'\vert \simeq 
ik R - i {\vec k}{\vec z}~'$, 
where $R=\vert {\vec x} \vert$ is the distance from the center of the
shower to the observation point. It corresponds
to the condition that observation distance $R$ exceeds the Fresnel 
distance $R_F=\pi n ~\nu ~ (L_s \sin\theta/2)^2 /c$, 
where $L_s$ is the typical length of the shower.
In this limit it is straightforward to show that the electric field
emitted by a shower in the 1-D approximation can be related to 
the Fourier transform of the longitudinal charge distribution:

\begin{equation}
\vec E(\omega,{\vec {\rm x}}) =
{e \mu_{\rm r}~\over 2 \pi \epsilon_0 {\rm c}^2}~i \omega
~\sin \theta~{{\rm e}^{ikR}\over R}~{\hat n_\perp}
\int dz'~Q(z')~{\rm e}^{i p z'}
\label{fraunhofer}
\end{equation}
where we have introduced for convenience the parameter
$p(\theta,\omega)= (1-n \cos \theta)~\omega /c$ in Eq.~\ref{fraunhofer} to
stress the connection between the radio emission spectrum and the
Fourier transform of the (excess) charge distribution.
This allows a simple analogy to the classical diffraction pattern of
an aperture function and helps understanding many of the complex
features of the results obtained by simulation.

For the case of a single particle moving between two fixed points
this expression (replacing $c$ by an arbitrary particle velocity $v$) 
reproduces the formula obtained in \cite{zas92}: 
\begin{equation}
{\vec E}(\omega,{\vec {\rm x}})=
{e \mu_{\rm r}~i \omega \over 2 \pi \epsilon_0 {\rm c}^2}~
{{\rm e}^{i k R } \over R} ~  
{\vec v}_{\perp}~ 
\left[{{\rm e}^{i(\omega - \vec k \cdot \vec v) {\rm t}_2} -
      {\rm e}^{i(\omega - \vec k \cdot \vec v) {\rm t}_1} 
                        \over i (\omega - \vec k \cdot \vec v)} \right]
\label{t1t2}
\end{equation}
where $\vec v_{\perp}$ refers to the particle's velocity projected in 
a plane perpendicular to the observing direction and $\rm t_2$ $(\rm t_1)$
is the time corresponding to the final (initial) point of the track. 
This is the basic expression used for the numerical simulation of
radio pulses from individual tracks (see appendix A).  


\section{The one-dimensional approach}

We will firstly explore the validity of the 1-D approximation by direct 
comparison with simulation results in three dimensions. The  
program we use for the full simulation of electromagnetic showers
in homogeneous ice, is described in Ref.~\cite{zas92}. 
The results of the simulation 
will be compared to those obtained using Eq.~\ref{fraunhofer} with 
different curves for the excess charge development function $Q(z)$  
what will turn out to be quite illustrative.  

\begin{figure}[hbt]
\centering
\mbox{\epsfig{figure=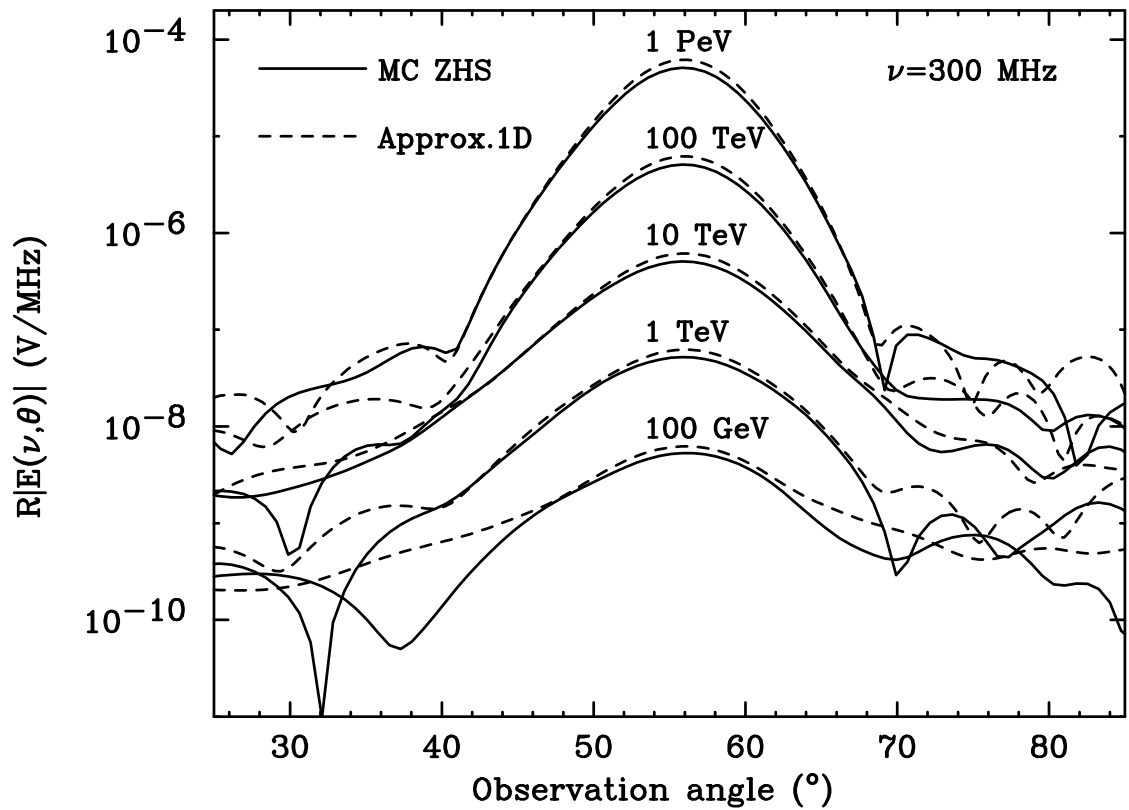,width=15.0cm}}
\vskip 1cm
\caption{Comparison of results of the 1-D approximation to 
fully simulated pulses for electromagnetic showers of 100~GeV, 
1~TeV, 10~TeV, 100~TeV abd 1~PeV. 
Simulations have been followed to threshold energy $E_{th}=1~$MeV. 
Shown is the angular distribution of the electric field amplitude 
for 300~MHz in the Fraunhofer limit multiplied by observation distance.}
\label{fig1}
\end{figure}

\begin{figure}[hbt]
\centering
\mbox{\epsfig{figure=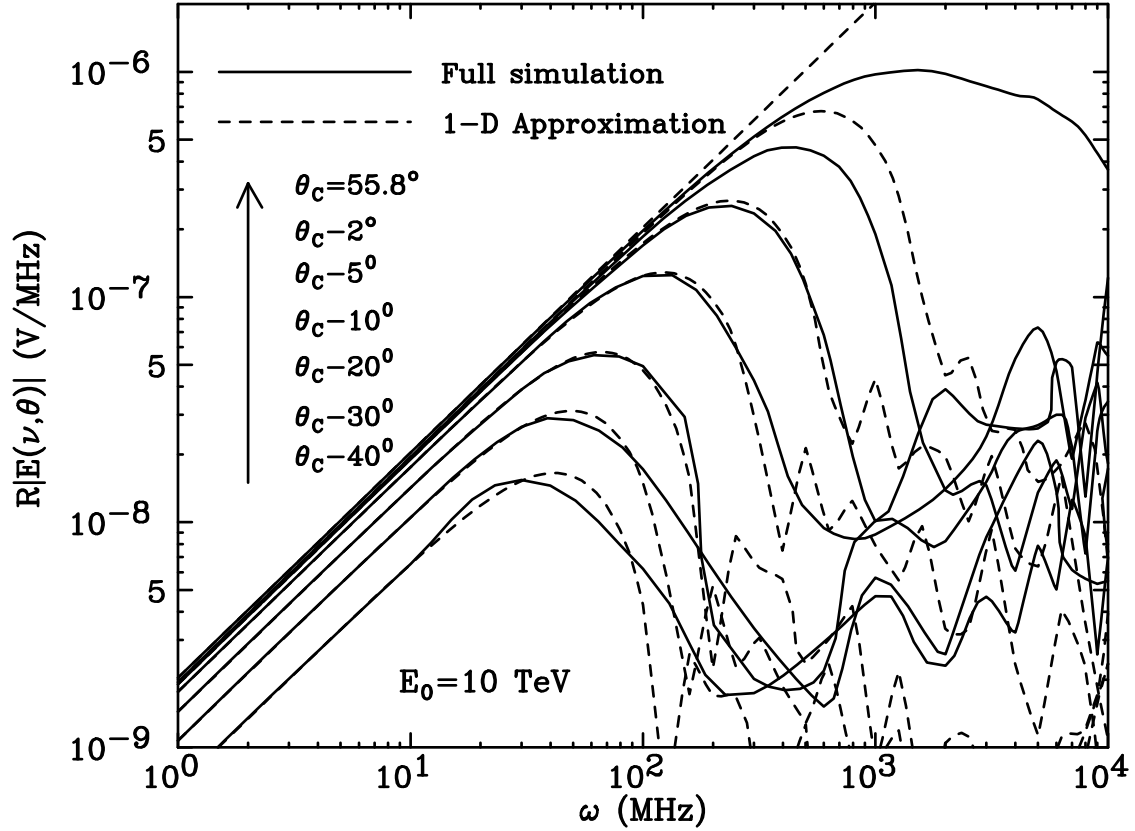,width=15.0cm}}
\vskip 1cm
\caption{Comparison of results of the 1-D approximation to 
fully simulated pulses for a 10~TeV shower with $E_{th}=611~$keV. 
Shown is the frequency spectrum of the electric field amplitude 
for different observation angles.}
\label{fig2}
\end{figure}

Fig.~\ref{fig1} compares the angular distributions of the pulses 
for showers initiated by different energy electrons using 
the full simulation and using Eq.~\ref{fraunhofer} with $Q(z)$ directly 
from the excess charge depth distribution as obtained in the same 
simulations. Fig.~\ref{fig2} displays the frequency spectra at different 
observation angles for a 10~TeV shower again for both approaches. 
Several conclusions can be drawn from these graphs with respect 
to the validity of the 1-D approximation. Clearly the electric field 
amplitude around the \v Cerenkov cone is well reproduced in shape by 
the 1-D approximation except in the \v Cerenkov direction where the 
approximation overestimates the amplitude by a factor that increases 
with frequency. Below 100~MHz the effect is negligible becoming of order 
$20\%$ (a factor of 2) for 300~MHz (1~GHz). 
The angular interval over which the approximation is valid slowly 
increases with shower energy and scales with the inverse of the frequency. 
Well outside the \v Cerenkov cone no agreement can be claimed 
but the order of magnitude of the approximation agrees with the simulation. 

For completitude we give a new parameterization for the frequency spectrum
in the \v Cerenkov direction using a finer subdivision of individual
electron tracks (approximation $a$ see appendix A),
which represents slight increase at frequencies above 500~MHz from that
given in Ref.~\cite{zas92}:
\begin{equation}
R \vert {\vec E}(\omega,R,\theta_C) \vert\simeq 2.53 \times 10^{-7}
\left[{E_{\rm em} \over 1~{\rm TeV}}\right]~
\left[\nu \over \nu_0 \right]~
\left[1 \over 1 + (\nu / \nu_0)^{1.44} \right]~{\rm V~MHz}^{-1}
\label{parameterization}
\end{equation}
where $\nu_0=1.15$ GHz.
This parameterization is valid to frequencies below $\sim 5~$GHz.

It is worth discussing the interpretation of the behavior of this 
approximation before we attempt to understand its validity in 
more complicated showers such as those having strong 
LPM effects \cite{LPMStanev,Klein}. 
In the \v Cerenkov direction, corresponding to $p=0$, 
the agreement between the approximation and the full simulation 
is excellent for frequencies below about 100~MHz. 
This corresponds to complete constructive interference characterized 
by a spectrum that increases linearly with frequency as shown in 
Fig.~\ref{fig2}. 
Above 100~MHz the simulated frequency spectrum deviates from linear behavior 
because the wavelength becomes comparable to the transverse deviation 
of shower particles \cite{zas92,tesis} and to a lesser extent because 
of time delays \footnote{It has been checked by direct simulation
that the time delays only become important 
for frequencies in the 10 GHz range
at the \v Cerenkov direction.}. 
Both these effects are ignored in the 1-D approximation 
that keeps on rising linearly. 
Away from the \v Cerenkov cone the approximation becomes valid even
to higher frequencies. This is because destructive interference is 
in this case due to the longitudinal excess charge distribution which 
is correctly taken into account by the approximation.

In spite of the approximation overestimating the amplitude of the 
electric field in the \v Cerenkov direction for frequencies above 
$\sim 100~$MHz, an ad-hoc correction can be implemented based on 
the shape of the frequency 
spectrum as obtained in the simulations. Since this effect is due to 
the lateral distribution of the electromagnetic component of the 
showers, it can be corrected with a unique function for each frequency. 
We have explicitly checked that the lateral distribution of 
electromagnetic showers is similar for showers with and without the 
LPM effect \cite{alz97,zalshowers}. 

We have calculated the difference between the 1-D approximation and 
the full simulation in the \v Cerenkov direction as a function of 
frequency what is shown in Fig.~\ref{diff-freq} for two different 
shower energies. Note that the difference is (up to a factor that scales 
with shower energy) the same for showers of different energies. 
For this calculation we have actually improved the simulation by 
splitting the individual tracks in small subintervals (approximation $c$, 
see appendix A). 
Also shown is the calculation without track subdivisions (approximation $a$)
for comparison. 
The angular behavior of the correction at a particular frequency 
can be also shown to be fairly independent of energy. 

The needed correction basically consists of rescaling the pulse just in 
the region around the \v Cerenkov direction. It can be achieved for 
instance dividing
the result of Eq.~\ref{fraunhofer} by a gaussian correction factor:
\begin{equation}
1+\left[{1D-FS \over FS}\right]~
{\rm e}^{-{1\over 2}\left[{\theta - \theta_C \over \sigma_{\theta}} \right]^2}
\label{correctionfactor}
\end{equation} 
The expression in brackets symbolically represents the relative difference 
between the frequency spectra as given by the 1-D approximation ($1D$)
and the full simulation ($FS$) calculated in the \v Cerenkov direction.
It simply sets the scale of the correction. The numerator is shown in 
Fig.~\ref{diff-freq} for two test cases, showing that it also scales 
with energy at least in the energy interval checked. 
The half width of the gaussian term is approximately given by: 
\begin{equation} 
\sigma_{\theta}=2.2^{\rm o} \left[{1~{\rm GHz}\over \nu}\right]  
\end{equation} 
For frequencies above the 100~MHz scale and high energies when the 
full simulation is not viable, one would implement the correction 
taking Eq.~\ref{parameterization} instead of the full simulated 
result. 

The 1-D approximation also works for 
complicated showers such as those initiated by electrons and photons 
of EeV energies with strong LPM effects \cite{alz97}. This has been 
explicitly checked by artificially lowering $E_{\rm LPM}$, the onset energy 
for LPM effects, so that showers 
with energies that allow full three dimensional simulations 
display the characteristic LPM elongations \cite{tesis}. 
The agreement between the full simulation and the 
1-D approximations is illustrated in Fig.~\ref{LPM} and it is 
clear that it is not limited to the central peak but also 
applies to the secondary peaks that appear in the angular distribution 
of the radiated pulse. The above correction prescription also works for 
these fictious elongated showers with a mild reduction in precision. 

Lastly, the simulation of the excess charge in 
an EeV shower can also be extremely time consuming because particles 
have to be followed at least to MeV energies when the interactions 
responsible for the excess charge become dominant over pair production 
and bremsstrahlung \cite{zas92}. 
According to simulations the pulse scales with the excess tracklength 
and this is practically only due to an excess of MeV electrons.
The excess number of electrons can be approximately obtained by rescaling 
the total number of electrons and positrons in a shower by 
the fraction of excess and total tracklengths. 
This factor is very stable and has a value of $25\%$ in ice \cite{zas92}
\footnote{This value corrects the previous conservative estimates
used in \cite{alz99} that quoted instead the ratio of excess projected 
tracklength to total tracklength as the relevant number (21$\%$).}.  
As convenient parameterizations of
the number of electrons and photons in showers are readily available
it is possible to calculate shower size distributions for very large
showers using them \cite{alz97,alz98}. 
In spite of the small 
gradual rise in the excess charge as the shower develops shown by 
simulations \cite{zas92}, the effects of this 
approximation are mild, a slight narrowing of the pulse which is 
negligible compared to the other approximations made (see 
Fig.~\ref{percent25}).  

Finally it is remarkably fortunate that neglecting lateral distributions
and time delays is a very good way of approaching the problem if
some considerations are cautiously taken into account, namely:

\begin{itemize}
\item{Take the Fourier transform of the longitudinal distribution of the 
excess charge $Q(z)$ (or one fourth of the total number of electrons and 
positrons if $Q(z)$ is not available) as given by Eq.~\ref{fraunhofer}.}
\item{For frequencies above 100~MHz divide the 1-D approximation by a 
correction factor as indicated by Eq.~\ref{correctionfactor} taking 
Eq.~\ref{parameterization} instead of the full simulation ($FS$) value.} 
\end{itemize}

\subsection{Discussion: The relation between radio pulses 
and shower fluctuations}

In the 1-D approximation the Fourier transform for $p=0$ becomes 
the integral of $Q(z)$, i.e. the excess tracklength. 
Simulations have shown that the excess tracklength scales extremely 
well with electromagnetic energy in the shower ($E_{\rm em}$) 
for both electromagnetic 
and hadronic showers up to energies exceeding 100 EeV with small 
fluctuations: 
\begin{equation}
t=6400 \left[{E_{\rm em} \over 1~{\rm TeV}}\right]~{\rm m}
\label{tracklength}
\end{equation}
Incidentally this nice property of the excess charge together with 
the fact that the radioemission in the \v Cerenkov direction is 
proportional to 
the excess tracklength, make such measurements excellent candidates 
for electromagnetic energy estimators. 
The breaking of the approximation at high frequencies is telling us
that the lateral distribution is playing
a significant role.

A simple limit of the 1-D approximation is obtained by taking an 
analytical expression for $Q(z)$ such as Greisen's parameterization for 
the average development of an electromagnetic shower \cite{greisen}. 
The result just gives the radiation in the \v Cerenkov cone but no 
radiation outside, just like the Fourier transform of a gaussian.

Invoking superposition we can subtract from a given shower development curve 
a smooth Greisen-like curve having the same tracklength. The result
displays the ``roughness" of the depth development curve
and we shall refer to it as the {\sl difference function}.  
The electromagnetic pulse is the sum of an isolated \v Cerenkov peak 
due to the Greisen-like curve and an extra contribution from the 
Fourier spectrum of the difference function which precisely vanishes at 
the \v Cerenkov direction because it does not contribute to the total 
tracklength. Moreoever, for ordinary showers 
the amplitude of the difference function becomes 
smaller relative to shower size as the shower energy increases. 
This is just an statistical effect of having a larger number of 
particles and it indicates that the ``spatial correlations" 
\footnote{The name stresses the fact that they are 
different from standard fluctuations in shower theory because they 
refer to variations in shower size for the same shower at different 
positions rather than comparing shower size at 
the same spatial position for different showers.}
contained in the difference function 
must be related to fluctuations in shower size. 

The fact that the magnitude of the difference function becomes smaller 
relative to shower size as the shower energy $E_0$ increases has the 
effect of "illuminating" the \v Cerenkov cone much more sharply 
with respect to directions well outside the \v Cerenkov direction. 
This effect can also be understood in terms of coherence. 
In the \v Cerenkov direction of greatest coherence the electric field 
amplitude scales with the shower energy $E_0$, 
but when the radiation is incoherent, i.e. well outside the \v Cerenkov 
direction, the electric field should add incoherently and hence scale 
with $\sqrt{E_0}$. This roughly agrees with simulations and  
nicely connects the properties of the radio emission to 
spatial correlations in shower development. 

  
For LPM showers the structure of the pulse outside the central 
narrow peak is still 
dominated by the longitudinal development of these showers because 
the amplitude of the difference function is much larger than for a 
conventional shower. This is because LPM showers fluctuate a great deal. 
(One can picture a characteristic LPM shower as a 
superposition of smaller subshowers with typical smooth profiles 
with random starting points along the shower length.) 

In other words, the Fourier modes of the excess charge distribution 
are probed by the electric field at a given value of $p$ and hence
at a given value of $\theta$ for a fixed frequency. The scale of
the correlations in the distribution (the ``wavelength" of the
corresponding mode) is inversely
proportional to $p$. As long as the scale of these correlations
is larger than the characteristic
lateral structure of the shower, the 1-D approximation
is expected to work. This is precisely what happens for the 
LPM fluctuations. 

In summary there are two angular regions for the electromagnetic 
pulse with a not very well defined boundary. One angular region 
corresponds to 
the surroundings of the \v Cerenkov cone where the 1-D 
approximation has powerful predictive power 
when one accounts for the correction described above. There 
is another region well outside the \v Cerenkov cone in 
which the calculated electric field amplitude drops considerably and 
behaves erratically, as some kind of ``white noise" 
corresponding to the incoherent regime. 
In this region the short scale correlations 
of the excess track distribution are being probed and here 
the predictive power is lost with the approximations discussed. 
To calculate the radiopulse in such regions one needs three 
dimensional simulation programs which must sample tracks in small
subintervals and which must follow all particles to the 100 keV 
region (approximation $c$ described in appendix A). 
All these requirements make it impossible with current 
computing power to simulate beyond 100~TeV. 
However the region outside the \v Cerenkov cone, 
having much reduced radioemission, is not 
very relevant for shower detection. 

\section{The validity of the Fraunhofer approximation}

All the calculations made of radio pulses have been made in 
the Fraunhofer approximation which corresponds to the limit:  
\begin{equation}
R > R_F  
= 3~{\rm m}~\left[ {{L_s} \over 1~{\rm m}} \right]^2 
\left[{\nu \over 1~{\rm GHz}} \right]   
\label{Frauncondition}
\end{equation}
%
Taking $L_s\sim 3.8~$m, 
corresponding to the nominal ten radiation lengths in ice 
of a electromagnetic (hadronic) shower below about 10 PeV (10 EeV), 
a frequency of 1~GHz and $\theta$ equal to 
the \v Cerenkov angle then $R_F\sim 45~$m. This distance
is to be compared with 
the km scale set by the small absorption coefficient of radio waves 
in cold ice which tells us 
that the Fraunhofer condition is clearly satisfied. 
For very long showers (such as those that display a 
very strong LPM effect) and high frequencies, $R_F$ 
exceeds the typical attenuation 
scale. As the distance $R$ is reduced 
to values below $R_F$, the diffraction pattern 
gradually turns into a Fresnel pattern in which the angular 
features  become blurred. 

It is possible to calculate diffraction patterns for such showers 
with the typical restrictions that apply to these simulations. 
A full calculation is again not viable for the shower energies 
at which this effect becomes important at km scale distances. 
We have calculated the radio pulses as observed at distances in 
which the Fraunhoffer approximation breaks down, using simulated 
electron showers of different energies. 
We apply Eq.~\ref{fresnel} for calculating electric field 
amplitudes at distances of order the Fresnel distance $R_F$, 
(a one dimensional transform that does not take the Fraunhofer 
limit). We calculate the effects for a range of energies and 
observation distances to specify 
the conditions under which the properties of the emission in the 
Fraunhofer limit are still valid. 

In Fig.~\ref{fresnel1EeV} we display 
the \v Cerenkov peak structure at 100~MHz for a range of distances around 
the Fraunhofer limit for a 1~EeV electromagnetic 
shower spanning 135 radiation lengths. 
We define the distance in relation to the center of charge of the shower. 
The calculated pattern has a reduced amplitude at the peak and becomes 
broader as expected. 
The Fraunhofer approximation is good to better than $10\%$ in absolute value 
for distances above $\sim 400~$m. 
For a 100~EeV (10~PeV) shower the distance increases to 5~km 
(decreases to 20~m) for a roughly similar accuracy. 
The angular width of the pulse in the near field case increases 
with respect to the Fraunhofer case roughly by $20\%$ when the 
amplitude reduces by $10\%$. 

In Fig.~\ref{ratios} we plot the ratio of the calculated and Fraunhofer 
amplitudes at the \v Cerenkov peak as a function of distance to the 
shower for different frequencies and shower energies. 
Also indicated are the absorption lengths at three different 
temperatures for reference. This graph sumarizes the results, for 
1~km distance and energies above a few hundred PeV, Fresnel effects will 
become a serious concern for GHz frequencies. 
Provided that the distance to the 
shower and its direction can be determined, Fresnel effects 
could be corrected for, but this would clearly complicate and limit the 
analysis. This suggests that lower frequencies in the 100~MHz or even below 
may be appropriate for EeV showers.
For hadronic type showers however no effects are foreseen
for energies up to the 10 EeV range except for few abnormally long
showers that are unlikely to happen \cite{alz98}. 

\section{Summary and Conclusions} 

We have shown that the calculation of coherent \v Cerenkov 
radio pulses from high energy showers in ice in the Fraunhoffer limit 
can be well approximated by neglecting the lateral distributions 
of the particles assuming that they travel at constant speed ($c$). 
The electric field amplitude simply becomes the one 
dimensional Fourier transform of the excess charge depth distribution. 
For the most relevant region around the \v Cerenkov 
direction, the approximation is correct for frequencies below 100~MHz. 
At higher frequencies the approximation is still relatively good 
but systematically overestimates the pulse in the \v Cerenkov direction. 
We have shown that the model can be made to agree at least up to 
1~GHz by subtracting a simple ad-hoc gaussian correction that is 
proportional to the shower energy and otherwise only dependent on 
frequency. We have reported the relevant parameters for the correction and 
have presented an improved parameterization for the electric field amplitude 
in the \v Cerenkov direction. 
 
We have also shown that instead of the actual charge excess distribution 
one can use the shower size longitudinal development curve which is more 
conventional than the excess charge, scaling the amplitude of the central 
peak by the excess tracklength fraction $0.25$. 

We have developped a similar approximation for the region in 
which the Fraunhofer limit ceases to be valid. 
We have finally studied the behavior of the radiopulses 
of long electromagnetic showers in this region. 
Our results are again suggesting to use low frequencies for 
EeV showers as concluded in Ref\cite{alz99}. These frequencies have 
a number of advantages because they are less attenuated, they allow  
observation of the angular structure with less detectors, 
and they have milder Fresnel effects at a given distance. Because 
of Fresnel corrections, the 
possibility of extracting the mixed character of electron neutrino 
interactions suggested in \cite{alz99} requires frequencies below 
100~MHz if the electron initiated subshower exceeds about 10~EeV.  

Lowering the frequency implies a higher threshold for detection 
because the \v Cerenkov spectrum increases with frequency but for EeV 
showers this should not be a problem. It has been estimated that the 
threshold for detecting showers at 1~km distance with 1~GHz broadband 
antennas is in the 10~PeV range \cite{zas92}. 
Since the signal to noise roughly scales with the square 
root of the bandwith which directly relates to the central frequency, 
a factor of 100 reduction in frequency will only call for about a factor 
of 10 enhancement of the threshold still giving a very large signal 
to noise ratio for EeV showers. 

Although our tests of the 1-D approximation rely heavily on a 
specific simulation program \cite{zas92}, our claim on the validity
of the 1-D approximation is model independent. For testing purposes
we used the 
charge excess distribution and the emitted radiopulses as obtained 
by the same routine. Numerically our results only apply for 
ice but it is only natural to expect that the same procedures can be 
applied to calculate the radiation in other materials. 

\vskip 0.5 cm
{\bf Acknowledgements:} We thank P.~Gorham for many early discussions 
about Fresnel corrections and D.~Besson, D.W. McKay, 
J.P. Ralston, S. Razzaque, D. Seckel and S. Seunarine for constructive 
criticism of the Montecarlo and many discussions. This work was supported 
in part by CICYT (AEN99-0589-C02-02) and by Xunta de Galicia (XUGA-20602B98). 
J. A. thanks the Department of Physics, University of 
Wisconsin, Madison and the Fundaci\'on Caixa Galicia for financial 
support. E. Z. thanks the Department of Physics, University of
Wisconsin, Madison, where this work was finished for its hospitality, 
and the Xunta de Galicia for partially supporting this trip. 

\begin{center}
\bf{APPENDIX A: The ZHS Montecarlo}
\end{center}

The simulation program used described in \cite{zas92} 
is a specifically deviced program for calculating 
radio-pulses from electromagnetic showers that follows 
particles to $\sim 100$~keV, taking into account 
low energy processes and timing. 
The depth development results have 
been compared to analytical parameterizations
given in the Particle Data Book \cite{pdb}, with 
which they agree to a few percent.
    
The calculation of the radio emission uses 
Eq.~\ref{t1t2} 
for electron and positron tracks. 
Several approximations can be made according 
to different choices in the subdivision of the 
individual charged particle tracks. 
In Ref.~\cite{radiorome} three different choices, 
named approximations $a$, $b$, and $c$ have been 
compared, testing for convergence as the subtracks 
become smaller. 

\begin{figure}[hbt]
\centering
\mbox{\epsfig{figure=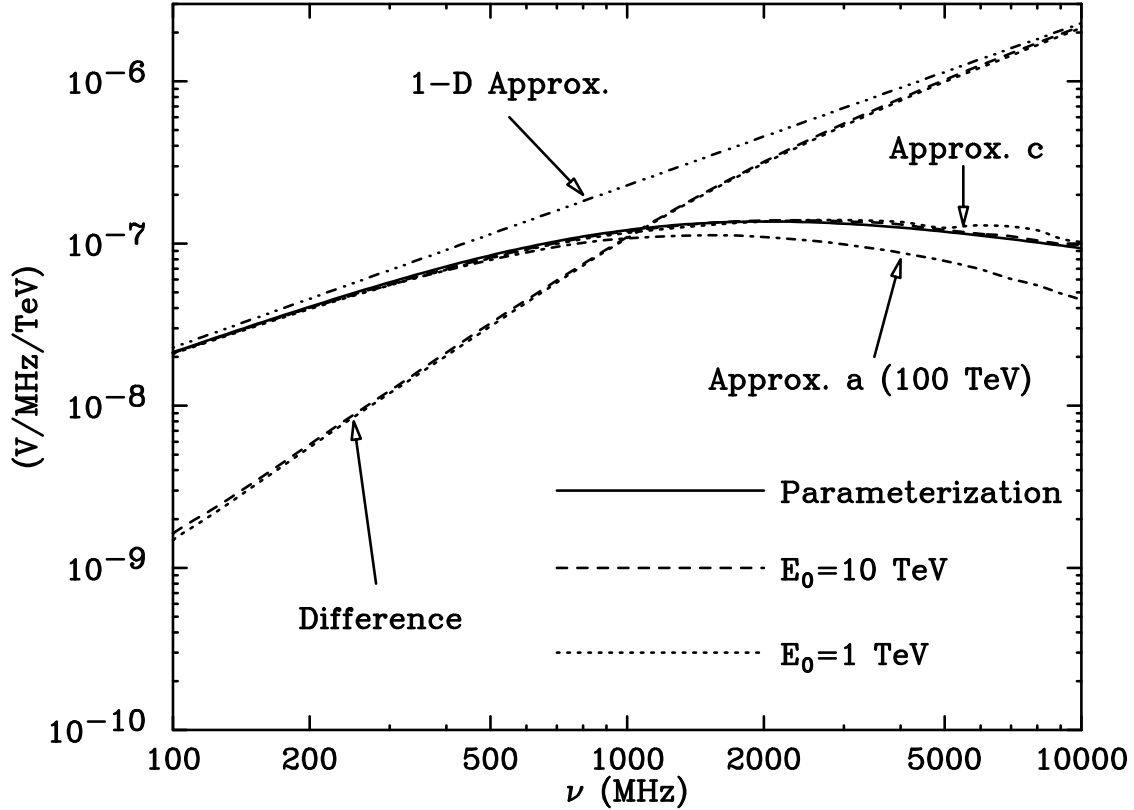,width=15.0cm}}
\vskip 1cm
\caption{Full simulation results for the frequency spectrum in the 
\v Cerenkov direction for 1 and 10~TeV electromagnetic showers in 
approximation $c$ and for a 100~TeV in the standard approximation 
($a$) used throughout for comparisons (see appendix A). 
They are compared to the results using the 1-D approximation (top curve). 
The improved parameterization for the $c$ approximation 
given by Eq.~(6) is also shown. 
The lower curves represent the difference between the 1-D approximation 
and the full simulation results (using approximation $c$). 
Note that both the spectrum and the difference have the same behavior 
for all shower energies.
All radiopulses scale with shower energy and are normalized to 1~TeV. 
}
\label{diff-freq}
\end{figure}

\begin{figure}[hbt]
\centering
\mbox{\epsfig{figure=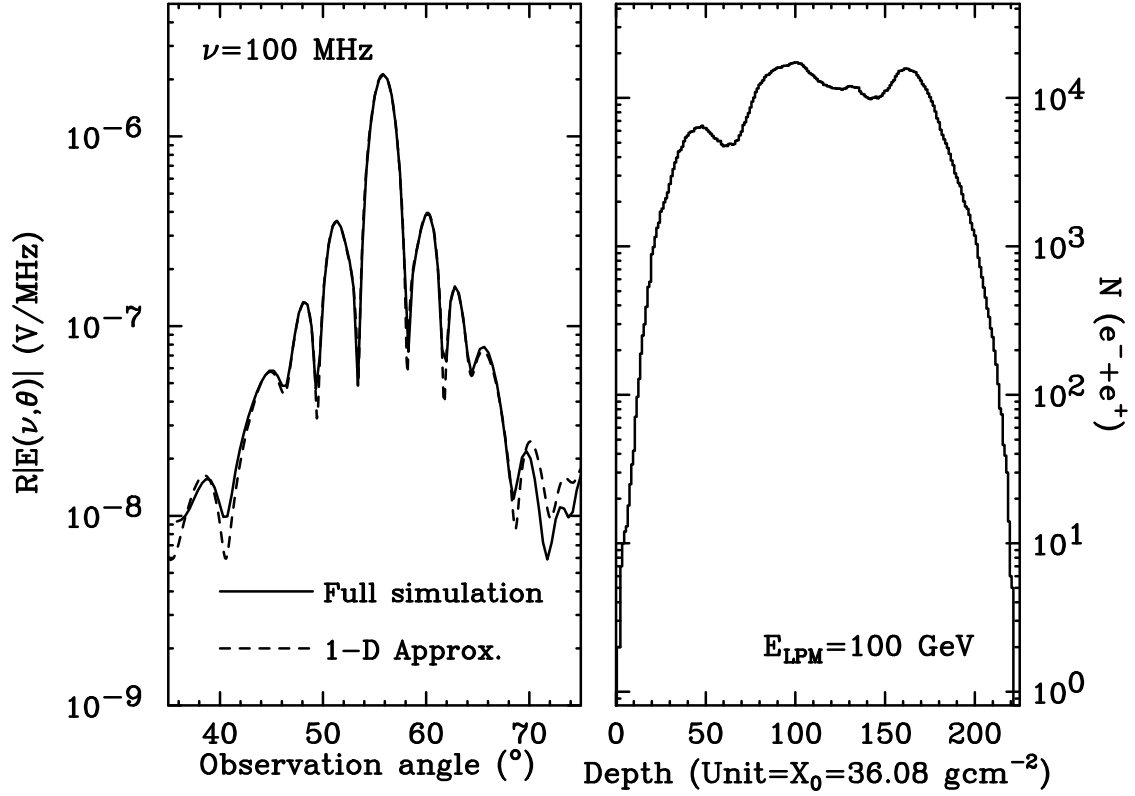,width=15.0cm}}
\vskip 1cm
\caption{Comparison of results of the 1-D approximation to 
fully simulated pulses for a fictitious composite shower that 
combines two 10~TeV subshowers initiated at the origin 
and one 100~TeV subshower starting at a depth of 25 radiation lengths. 
Furthermore these subshowers are artificially elongated by reducing 
the onset of the LPM effect ($E_{\rm LPM}$=100~GeV instead of the actual 
value for ice which is 2~PeV). The longitudinal shower profile is 
also shown.}
\label{LPM}
\end{figure}

\begin{figure}[hbt]
\centering
\mbox{\epsfig{figure=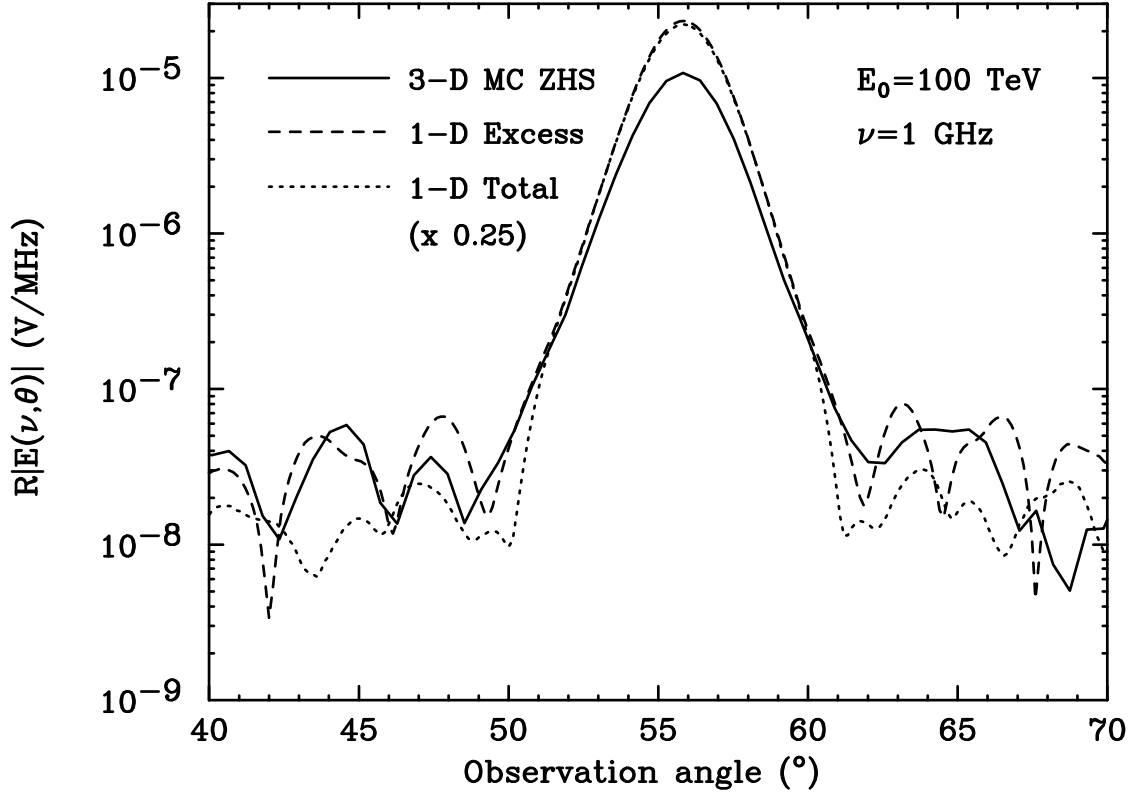,width=15.0cm}}
\vskip 1cm
\caption{Comparison of results of the 1-D approximation to 
a fully simulated pulse for an electron shower of 100~TeV. 
Simulations have been followed to threshold energy $E_{th}=611~$keV. 
Shown is the angular distribution of the electric field amplitude 
for 1~GHz in the Fraunhofer limit multiplied by observation distance. 
Two curves are shown for the 1-D approximation using the excess charge 
$Q(z)$ and the shower size $N(z)$ as obtained in the same simulation. 
The value obtained with shower size has been multiplied by $0.25$ as 
explained in the text.}
\label{percent25}
\end{figure}

\begin{figure}[hbt]
\centering
\mbox{\epsfig{figure=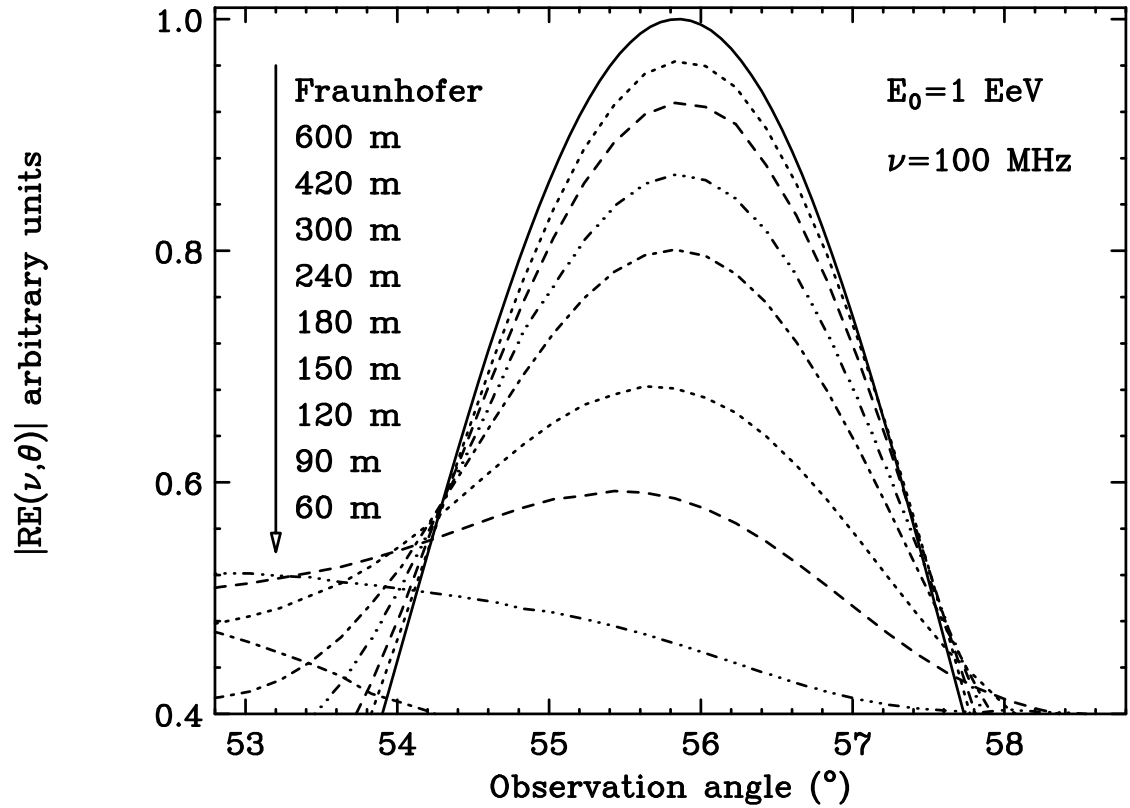,width=15.0cm}}
\vskip 1cm
\caption{Results of the 1-D approximation as the observation distance 
approaches the Fresnel distance $R_F$ for a 1~EeV electromagnetic 
spanning 135 radiation lengths.}
\label{fresnel1EeV}
\end{figure}

\begin{figure}[hbt]
\centering
\mbox{\epsfig{figure=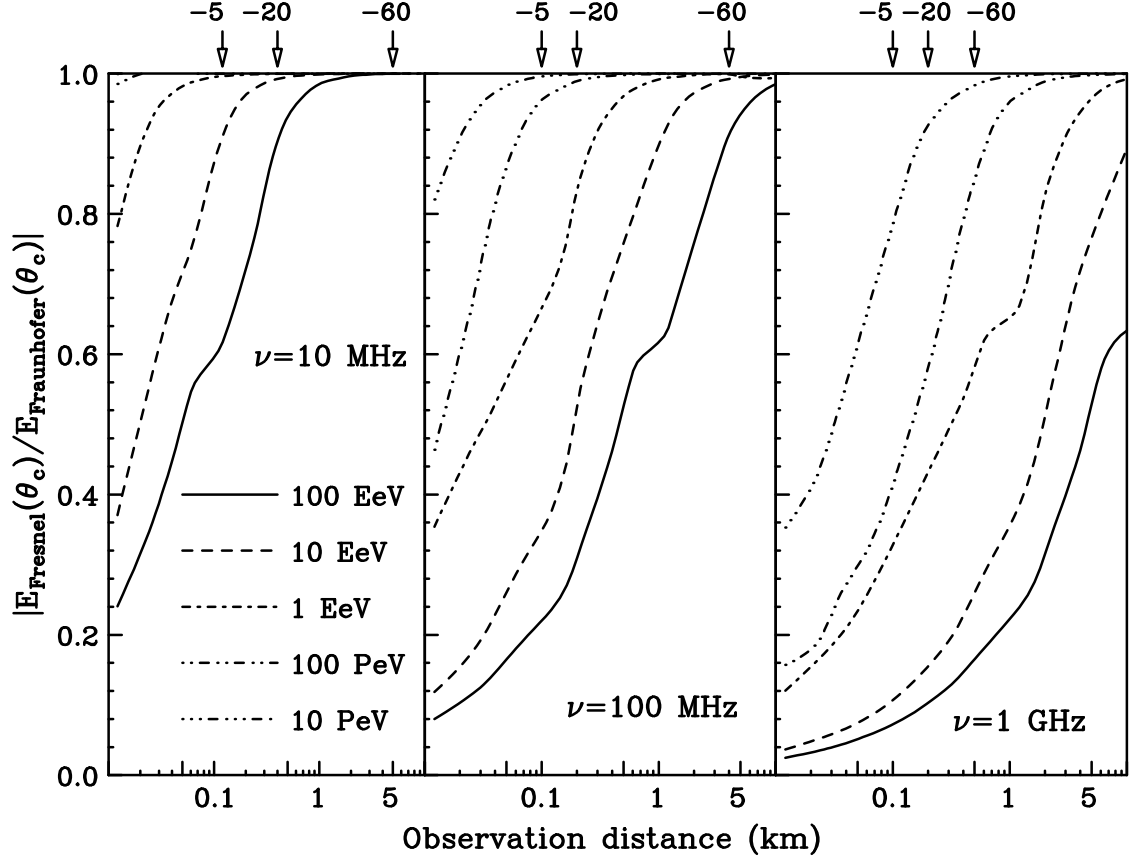,width=15.0cm}}
\vskip 1cm
\caption{Electric field amplitudes in the \v Cerenkov direction 
as a function of observation distance as obtained using the 
1-D approach for different frequencies. 
The amplitudes are normalized to the amplitude in the Fraunhofer limit. 
The arrows indicate the attenuation lengths for the corresponding 
frequencies at three different reference temperatures (in $^{\circ}$C).}
\label{ratios}
\end{figure}

Approximation $a$ is the standard 
that has been used in Refs.~\cite{zas92,zas91}. 
It corresponds to taking the end points of all 
the tracks, and it just uses the 
average velocity for the corresponding 
effective track in Eq.~\ref{t1t2}. 
This is the standard reference calculation 
used throughout in this article except for 
Fig.~\ref{diff-freq}. Note that this 
approximation gives the correct result provided 
the particle velocity is constant along the track. 

Approximation $b$ subdivides 
the electron tracks according the different 
interaction points found along the track, 
(multiscattering is not considered as an 
interaction here). This approximation 
subdivides the track in finer subintervals 
as the energy becomes smaller, because the 
low energy electron scattering cross sections 
exceed bremsstrahlung and pair production. 
For each subtrack the average velocity is 
calculated between the corresponding end 
points of the track. Finally approximation $c$ 
subdivides each interaction according to a 
convenient algorithm for spliting the 
propagation of particles designed to better 
calculate the multiple scattering at low energies.  

\begin{figure}[hbt]
\centering
\mbox{\epsfig{figure=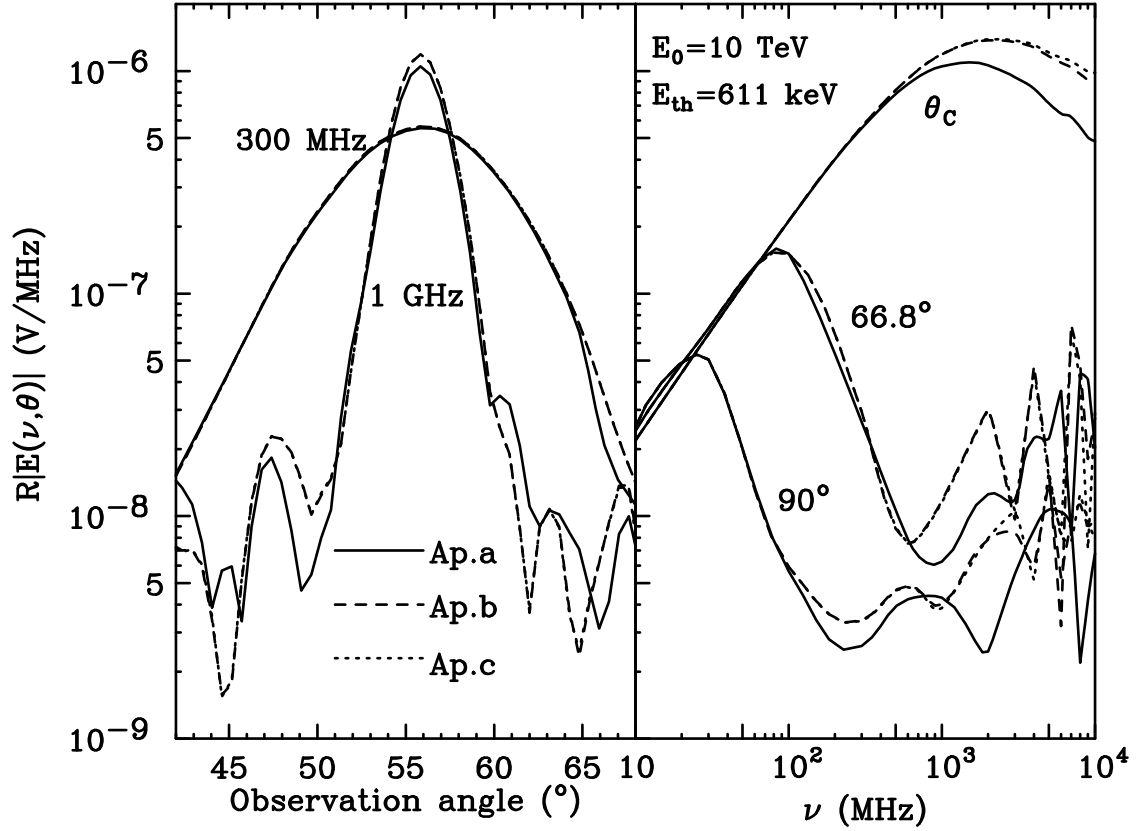,width=15.0cm}}
\vskip 1cm
\caption{Results of the full simulations with different algorithms 
for track subdivisions in the calculation of the electric field 
amplitudes as discussed in the text (appendix A). Shown are both the angular 
distributions for 300~MHz and 1~GHz and the frequency spectrum at 
three different observation angles for a 10~TeV electron shower 
with a threshold of 611~keV. }
\label{abc}
\end{figure}

The three approximations are compared in Fig.~\ref{abc} 
illustrating the convergence of the method and how 
the approximation $a$ is valid in the \v Cerenkov 
cone to a precision better than about $10\%$ for frequencies 
below 1~GHz. 
Full simulations in approximation $c$ are much more time 
consuming and have to be done for shower energies below 
$\sim 100~$TeV. At low energies fluctuations from shower to 
shower are more important so that these tests are inevitably 
subject to larger uncertainties because of such fluctuations. 

\begin{center}
\bf{APPENDIX B: The gaussian approximation}
\end{center}

For electromagnetic (hadronic) showers below 10~PeV (10~EeV), that 
is having no important deviations from Greisen behavior, the 
electric fied around the \v Cerenkov cone can be accurately determined 
with a gaussian approximation.
The precise width of the cone inversely relates to the width (in $z$)
of the excess charge depth distribution, $Q(z)$. 
As $p$ is directly related to the observation angle $\theta$ with  
an expression that involves the frequency as an overall factor, 
the width of the angular distribution of the ''central peak''
becomes inversely proportional to $\omega$. 

For small deviations from the
\v Cerenkov angle ($\Delta \theta$) the expression for $p$ to first order
is \cite{alz99}:
\begin{equation}
p = {\omega \over {\rm c}} \sqrt{n^2-1}~\Delta \theta+O(\Delta \theta^2)
\simeq 30.8 \left[ {\nu \over 1~{\rm GHz}} \right]~ \Delta \theta~
({\rm m}^{-1})
\label{papprox}
\end{equation}
The numerical value given in this expression corresponds to showers
in ice with $n=1.78$. 
Defining the gaussian width by the points in which the amplitude drops by
a factor $\sqrt{e}$ a gaussian of half-width $\sigma_z$ transforms to
another gaussian of half-width $\sigma_p = (\sigma_z)^{-1}$.
We can fit a gaussian to the excess charge depth development curve
identifying the {\sl shower length} by the width $l=2 \sigma_z$
and the angular full width of the radiopulse is then:
\begin{equation}
\sigma_{\theta} \simeq 3.72^\circ
\left[{1~{\rm GHz} \over \nu} \right]
\left[{1~{\rm m} \over l} \right]
\end{equation}
using approximation given by Eq.~~\ref{papprox}.
For a typical shower length of 8 radiation lengths
($\sim 3.1~$m in ice) the
angular width of the pulse is about $1^{\circ}$ at 1~GHz,
in agreement with Ref.~\cite{zas92}.

\newpage

\end{document}